# "Bubbles in Society"
# The Example of the United States Apollo Program

Monika Gisler[1] and Didier Sornette[2]

[1]ETH Zurich, D-ERDW, HPP L5, CH-8093 Zürich
monika.gisler@alumnibasel.ch
[2]ETH Zurich, D-MTEC, Chair of Entrepreneurial Risks, Kreuzplatz 5, CH-8032 Zurich
dsornette@ethz.ch

**Abstract** We present an analysis of the economic, political and social factors that underlay the Apollo program, one of the most exceptional and costly projects ever undertaken by the United States in peacetime that culminated in 1969 with the first human steps on the Moon. This study suggests that the Apollo program provides a vivid illustration of a societal bubble, defined as a collective over-enthusiasm as well as unreasonable investments and efforts, derived through excessive public and/or political expectations of positive outcomes associated with a general reduction of risk aversion. We show that economic, political and social factors weaved a network of reinforcing feedbacks that led to widespread over-enthusiasm and extraordinary commitment by individuals involved in the project as well as by politicians and by the public at large. We propose the general concept of "pro-bubbles", according to which bubbles are an unavoidable development in technological and social enterprise that benefits society by allowing exceptional niches of innovation to be explored.

**Keywords**, Risk, bubble, Apollo-Program, United States

## 1. Bad versus good bubbles: the "pro-bubble" hypothesis

In finance and economics, the term "bubble" refers to a situation in which excessive public expectations of future price increases cause prices to be temporarily elevated without justification from fundamental valuation. Thus, bubbles occurring in an economic context are seen as optimistic predictions about the future that prove wrong. In this respect, they are considered to be bad. Their ominous character is amplified by the uncertainty stemming from the lack of a consensus on their causes, which make them a major challenge to economic theory. Numerous books and scholarly works attempt to dissect the developments, origins and wealth destructions associated with financial bubbles (White and Rappoport, 1995; White, 1996; Galbraith, 1997; Kindleberger, 2000; Shiller, 2000; Shefrin, 2000; Shleifer, 2000; Sornette, 2003). Financial bubbles are also of great concern to central banks and regulatory agencies as they may signal and/or cause systemic instabilities (Davis, 1995).

In the present essay, we build on the somewhat antagonistic hypothesis proposed recently by Sornette (2008) that bubbles constitute an essential element in the maturation process and in the dynamics leading to great innovations or discoveries. Moreover, we contend that bubbles seem to be an unavoidable development in technological and social enterprise that benefits society by allowing exceptional niches of innovation to be explored. During bubbles, people take inordinate risks that would not otherwise be justified by standard cost-benefit and portfolio analysis. Bubbles are characterized by collective over-enthusiasm as well as unreasonable investments and efforts, derived through excessive public and/or political expectations of positive outcomes. Only during these times do people dare explore new opportunities, many of them unreasonable and hopeless, with rare emergences of lucky achievements. Sornette (2008) proposed this mechanism as the leading one controlling the appearance of disruptive innovations and major advances. Bubbles, defined as collective over-enthusiasm, seem a necessary (and unavoidable) evil to foster our collective attitude towards risk taking and break the stalemate of society resulting from its tendency towards stronger risk avoidance. An absence of bubble psychology would lead to stagnation and conservatism as no large risks are taken and, as a consequence, no large return can be accrued. We refer to this concept as the "pro-bubble hypothesis". The concept of the "new economy", a term used during the 1920s (the utility bubble), the 1960s (the "tronic" boom) and the 1990s (the Internet and Communication Technology bubble), captures vividly this mindset held by the large majority of investors and firm managers at times of bubbles: there is widespread belief that times have changed irreversibly and for the better, and that a new epoch, a new economy, a new prosperity without business cycles and recessions, is the novel emergent rule with the expectation of endless profits. Our goal is to explore and test the evidence that bubbles lead to a lot of destruction of value but also to the exploration and discovery of exceptional niches.

Our pro-bubble hypothesis is related to and extends a few previous works in the literature. In her book (Perez, 2002), Carlota Perez uses the term bubbles to describe the financial processes characterized by the installation of a new paradigm (or "revolution") and its concentration of investment in the respective new (scientific) enterprise (infrastructure, human resources, etc.). Her analysis suggests that the working of markets cannot by itself explain the recurrence of major crashes and depressions. Instead, the emergence of these phenomena need to be explained by the analysis of the tensions, resistance, obstacles and misalignments that arise from within the wider social and institutional scene. In a forthcoming paper, Perez (2008) develops further this argument by showing that the establishment of all major infrastructures associated with dominant techno-economic paradigms has been intimately linked to major technological bubbles, entailing the euphoric and reckless build-up of overcapacities of various kinds. Dosi und Lovallo (1997) furthermore suggest that cognitive biases may lead to collective social gains such as the collective value of overconfidence that often drives individual entrepreneurial decisions. Collective boom and bust behaviors drive private investors to develop externalities and collective physical infrastructures that no sober exclusively profit-motivated actor would have done otherwise (Gary, Dosi and Lovallo, 2007). Columnist Daniel Gross, in his 2007 book on *Why Bubbles Are Great For The Economy*, develops a similar argument. He takes a counterintuitive look at economic bubbles – those once-in-a-generation crazes that everyone knows can't last, and don't. Common thinking states that excessive investment in fixed assets is bad for investors, for the employees of the bubble companies, and for the economy. Gross surveys historical and modern bubbles and describes evidence for the benefits to be far more durable than the disruptions: in each case, most investors flopped, but businesses and consumers found themselves with a usable commercial infrastructure. He contends that the outcomes built during infrastructure bubbles do not get plowed under when their owners go bankrupt; on the contrary, they're getting reused later by those with new business plans, lower cost bases, and better capital structures. In fact, what is likely to be a catastrophe from one point of view, others, such as Perez (2002, 2008) and Gross (2007), see a collective social gain stemming from bubble behavior (major investments with low returns) on the long run. Innovation processes and the creation of new technology seem inherently associated with bubbles. These authors suggest that crashes/crises/busts are unavoidable epochs covering about 10% of the time, but on the other hand they provide benefit for the remaining 90% of the time.

Essentially, our pro-bubble hypothesis extends these previous works by emphasizing the dynamical role played by bubbles in reducing collective risk aversion that occur during the innovation and discovery processes. Over-optimistic expectations lead people to focus almost solely on the returns and to forget the risks. Returns are understood here not only as financial profits, but also as individual and/or social gains. The main questions resulting from this hypothesis are: What factors (economically, politically, socially) have been crucial in the discussed processes? What risks were undertaken and within which frame (temporal, cultural), by whom (individuals as well as institutions), and how? Which are the investments, what are the nature and level of efforts?

In order to learn from historical analysis and contribute to contemporary strategic and policy thinking, the scientific program suggested by the pro-bubble hypothesis is to identify the building of bubbles in scientific/social environments in different periods, to analyze their allegedly collapse and to identify the socio-economic processes taking place in the aftermath. Sornette (2008) suggested five examples illustrating the pro-bubble hypothesis: the great boom of railway shares in Britain in the 1840s, the Apollo program of the United States, the Human Genome project, the closing of mammals (Dolly, the sheep), and the ITC bubble culminating in 2000. Here, we study in depth one of them, the United States Apollo program, as it represents arguably one of the most striking modern examples of a fundamental innovation process. The program was launched in 1961 by then President J.F. Kennedy, being fulfilled by the *Apollo 11* lunar landing in 1969, and concluded with *Apollo 17* in 1972. Our attention is on the question of the undivided willingness to take large risks that give way to major innovations.

The organization of the paper is as follows. Section 2 presents a general introduction and historical overview of the Apollo program. Section 3 describes the development processes underlying the political and financial supports to the Apollo program. Section 4 explores its human and social dimensions, and documents the features that are characteristics of a bubble. It also contrasts the mindsets and behaviors of the protagonists and of the public before, during and after 1969. Section 5 concludes in particular by comparing the expectations developed during the Apollo bubble and what actually happened in the following decades.

## 2. The United State Apollo program
### 2.1 Overview

The Apollo program was one of the most ambitious and costly single project ever undertaken by the United States in peacetime. One of the main foci of scholars studying the Apollo program has been from the perspective of the space race between the United States and the USSR in the context of the cold war. The cold war and the space race were indeed important factors in the formation of the Apollo program; however, we argue that there were other equally important and perhaps even more important factors at play. Support for this hypothesis can be found in the fact that the Cold War did not end with the termination of the Apollo project; neither did collaboration between the US and the USSR start after the collapse of the latter in 1989, on the contrary, the SOJUS project started as early as 1967. Or, as John Logdson has emphasized, it was "[…] global space leadership that was the basic goal, with U.S.-Soviet space relations an important, but not the only, venue for achieving that leadership." (Logsdon, 2007: 90; see also Launius, 2006). Eisenhower's policy was to deny the existence of a space race between the United States and the USSR, and Kennedy first continued this policy by proposing to the Soviets to collaborate in space exploration (Etzioni, 1964, Diamond, 1964; Logsdon, 1979, 2007).

Our interest therefore focuses on the internal perspectives that played a role in the development of the Apollo program. It is important to note that the Apollo program enjoyed high visibility and strong interest from a large fraction of the population, including its financial and technical components. We argue that the Apollo program developed as a bubble, first nucleated and engineered by a special interest group, which inflated to a very large size only through the general positive feedback mechanisms discussed in the introduction that are usually associated with bubbles. One characteristic feature of a bubble is overinvestment (with respect to standard cost-benefit calculations) into new technologies and sciences. Here, we understand the phenomenon of overinvestment as occurring not only in the financial sphere, but also in the working commitments and loads that people were ready to undertake, the personal engagement for new ideas, the power of leadership that attracted potential sponsors as well as the public. We call this process an exogenous critical bubble (Sornette, 2005), the term exogenous emphasizing that the process was engineered, and the word critical stressing the fact that after some encouragement the program flew by its own in the public mind, leading to explore uncharted territories independent of the expected results. Indeed, as we recall below (and as is well-known), the Apollo program enjoyed a tremendous support, financially as well as societal. Yet, it ended rather abruptly. Looking from a different angle, we try to show though that the ending was not as unexpected as it seemed to have been, when viewed from the perspective offered by the pro-bubble hypothesis.

**2.2 Historical perspective**

In 1958, Wernher von Braun (1912–1977), one of the most important rocket scientists and defenders of space exploration during the period between the 1930s and the 1970s, announced the plan to land on the Moon within a decade. On May 25, 1961, John F. Kennedy, who had succeeded Dwight D. Eisenhower as the US-President a few months before, challenged his nation to land a man on the Moon before the decade was out ("... this nation should commit itself to achieving the goal, before this decade is out, of landing a man on the Moon and returning him safely to the Earth.") (Congressional Record (25 May 1961) vol. 107, pt. 7: 8881). On April 24, at a meeting of the Space Exploration Program council, Vice President Lyndon Johnson furthermore asked: "We've got a terribly important decision to make. Shall we put a man on the Moon?" Everybody agreed (Orloff and Harland, 2006; Bizony, 2006). By July 20, with hardly a dissenting vote, Congress authorized a space budget 60 percent higher than Eisenhower's January request (Fortune, Nov. 1963). And in fact – as we all know –, on July 16, 1969, the spacecraft *Apollo 11* took off from Cape Kennedy (now Cape Canaveral) in Florida, sending three American astronauts into space. Three days later, Neil Armstrong (1930–) stepped off the missile and walked on the Moon's surface, his colleague Edwin E. Aldrin Jr. (1930–) to follow eighteen minutes later. Project Apollo (at that time under the presidency of Richard Nixon), which had been born out of the Mercury Project that successfully sent manned capsules into orbit, proved to be one of the most successful endeavors in the history of the National Aeronautics and Space Administration (NASA) (http://www.astronautix.com/project/apollo.htm).[1]

Answering von Braun's dreams and President Kennedy's challenge of landing men on the Moon by 1969, the Apollo program was launched, America's premier space effort during the 1960's, and the centerpiece of America's manned space effort (Ertel and Morse, 1969–1978; Bilstein, 1994; Nagel and Hermsen, 2005; Orloff and Harland, 2006; Becker, 2007; http://www.nasa.gov/topics/history/index.html [URL: April 30, 2008]). It was considered a highly important project for its focus on innovation and the requirement of the most sudden burst of technological creativity, enormous human efforts and expenditures, and the largest commitment of resources (estimated variously between roughly 21.8 and 25 billion United States Dollars (USD) – 1960s value).[2] At its peak, the Apollo program employed 400,000 people, and required the support of hundreds of universities and 20,000 distinct industrial companies (Ertel and Morse, 1978). Enthusiasm was high in terms of funding as well as in terms of human effort. Its objective was twofold and its expectation high: the immediate goal, as proclaimed by President John F. Kennedy before Congress in 1961, was to land men on the Moon. A second and far broader objective was to make the United States preeminent in space, taking a leading role in space achievement and ensuring that the nation would be second to none in its ability to explore the expected riches of the solar system and beyond.

What made the people involved take risks to put machines and men in space? Three points are to be addressed: First, the development of science and technology was at the vanguard of these endeavors, fueled by scientific curiosity and ambition on the part of the scientists. Second, political reasons were equally strong: the US and the USSR were competing for military superiority as well as for dominance in space, the cold war was building up strongly. On October 4, 1957, the Soviet Union had successfully launched Sputnik I. The U.S. Defense Department responded to the political furor by approving funding for another U.S. satellite project. As a simultaneous alternative to Vanguard, one of the most important protagonists of the early space program, Wernher von Braun and his team began to work on the Explorer, later on the Saturn I project. After that, a state of affairs was launched what nowadays is referred to as the "space race." It involved the efforts to explore outer space with artificial satellites, to send humans into space, and to land people on the Moon. And third, space technology became a particularly important arena in this race. Space research had a dual purpose: it could serve peaceful ends, demonstrating a country's scientific competence but could also contribute to military goals.

## 3. Political and financial support to the Apollo program
### 3.1 Policies and Politicians

The Apollo program was originally conceived early in 1960, during the Eisenhower administration, as a follow-up to America's Mercury program (Callahan and Greenstein, 1997). The goal was to develop the basic technology for manned spaceflight and investigate human's ability to survive and perform in space. While the Mercury capsule could only support one astronaut on a limited Earth orbital mission, the Apollo spacecraft was intended to be able to carry three astronauts on a circumlunar flight and perhaps even on a lunar landing. While NASA went ahead with planning for Apollo, funding for the program was far from certain, particularly given Eisenhower's equivocal attitude to manned spaceflight. To answer concerns that the Americans were losing the lead, Eisenhower decided to create a civil establishment dedicated to space research, feeling that spending on space exploration could be seriously defended only in both military and scientific terms. On 29 July 1958, the creation of an American space agency, the National Aeronautics and Space Administration, NASA, was approved. "Space exploration holds the promise of adding importantly to our knowledge of the Earth, the solar system and the universe", claimed Eisenhower (D. Eisenhower cited in Cadbury, 2005: 187). More importantly, as a civilian agency, NASA would avoid service rivalries and satisfy political demand for peaceful uses of space (Dallek, 1997).

In November 1960, John F. Kennedy was elected President. As a senator running for president, he had argued in a questionnaire of the journal *Missiles and Rockets* that the United States were in a strategic space race with the Russians, and, what's more, that the Americans were allegedly losing this race (Kennedy, 1960; Diamond, 1964). It is with Kennedy then that the idea of space race against the Russians took its incipiency. Using space exploration as a symbol of national prestige, Kennedy initially warned of a missile gap between the two nations, pledging to make the United States the first nation in space (Beschloss, 1997). And it paid its tribute: In his autobiography, astronaut Michael Collins retrospectively expressed his concerns for the years 1962 and again 1965, the Russians being ahead in terms of space technology (Collins, 1974). By April 1963, however, Kennedy considered to justify the Apollo program in terms other than cold war prestige.

His maneuvering between personal belief and public statements was more than sheer rhetoric: Kennedy did not immediately come to a decision on the status of the Apollo program once he was elected President. He not only knew little about the technical details of the space program, what's more, he was put off by the massive financial commitment required by a manned Moon landing (Sidey, 1963). Despite public support, Kennedy had expressed concerns about the program and the funds that it absorbed. When NASA Administrator James Webb requested a thirty percent budget increase for his agency, Kennedy supported an acceleration of NASA's large booster program but yet deferred a decision on the broader issue. His plans were abruptly changed by two unexpected events in mid April 1961: The first man in Earth orbit by the soviets (Yuri Gagarin) and the CIA-backed Cuban exiles invasion failure at the Bay of Pigs (Benson and Faherty, 1978; Beschloss, 1997). In late May 1961, his budget director had warned Kennedy of the large price tag of Apollo and, when he met Nikita Khrushchev on Vienna the following month, Kennedy suggested that the United States and the Soviet Union explored the Moon as a joint project. The Soviet leader requested disarmament as a prerequisite for the US-USSR cooperation in space. In September 1963, Kennedy met with James Webb and told him to prepare for a joint program. On September 20, 1963, Kennedy gave his well-known speech before the United Nations General Assembly, in which he again proposed a joint human mission to the Moon. Publicly though, the Soviet Union was noncommittal (Benson and Faherty, 1978; Launius, 2007). Kennedy, as a result, seized on Apollo as the ideal focus for American efforts in space. He ensured continuing funding, shielding space spending from the 1963 tax cut and diverting money from other NASA projects. In the public, beside politics, Kennedy needed a different message to gain support. Even more so after Khrushchev dissociated his country from the space race.

His reluctance was also fueled by another apprehension: In 1961, he expressed great concern about the people involved: "We are, I hope, going to be able to carry out our efforts with due regard to the problem of the life of the men involved this year." (J. F. Kennedy cited in Dick and Cowing, 2004: 259). Even though he did not say it directly, he was referring to the risk of putting humans into space. James Webb, then NASA Administrator and the person who had most to gain from the tremendous expansions of budget and power, was the last to climb aboard for the Moon (Bizony, 2006). He issued a statement equally hesitantly: "We must keep the perspective that each flight is but one of many milestones we must pass. Some will completely succeed in every respect. Some partially, and some will fail." (J. Webb cited in Dick and Cowing, 2004: 259). The reward for the United States when people were willing to take risks and to explore through manned space flight was obvious to Kennedy; it was the public he had to make sure of it.

Later in 1963, Kennedy asked Vice President Johnson to investigate the possible technological and scientific benefits of a Moon mission (see the Memorandum for Vice President of April 20, 1961, in Launius, 2004). Johnson had no time for caution. He concluded that the benefits were limited but, with the help of scientists at NASA (and Webb had already assured him that a Moon landing was technically possible), he put together a powerful case, citing possible medical breakthroughs and interesting pictures of Earth from space. By emphasizing the scientific payoff and playing on fears of Soviet space dominance, Kennedy and Johnson managed to swing public opinion. The technology base that Apollo would enhance, so was the saying, was worth the risk – financially and personally. And the argument caught on (an example is given in Dick and Cowing, 2004).

Even more so, after the assassination of Kennedy: Johnson and Webb constantly defended the Apollo program as the wish of this slain president. That was a very powerful argument to be made in the political arena and they achieved success in protecting the program, even as everything else at NASA began to suffer budget cuts from the mid-1960s onward (Launius, 2007). After Johnson became President in 1963, his continuing defense of the program allowed it to succeed in 1969.

It was as Vice President though that Johnson entered the scene of the space program, going beyond what a Vice president normally did (Beschloss, 1997; Dallek, 1997). When Kennedy asked him to come up with arguments for space matters, Johnson turned to several individuals in order to gather information on the options and possibilities of a potential space program. Wernher von Braun, whom Johnson had contacted in April 1961, assured him that the Soviets would not have a rocket to send a crew to the Moon. Secretary of Labor Arthur Goldberg on the other hand disagreed with the premise that a Moon program would stimulate the economy as was anticipated by Kennedy.

Nonetheless, after having collected technical opinions, Johnson – from 1961 onwards being the chairman of the space council – began to persuade political leaders of the need for an aggressive lunar landing program. He spoke with several representatives to ascertain if they were willing to support an accelerated space program. Whenever he heard reservations, Johnson used his forceful personality to persuade the Kennedy administration.

It was clearly Johnson, even at an early state, who favored an expanded space program and a maximum effort to land an astronaut on the Moon. Johnson insisted that manned exploration of the Moon was essential whether or not the United States turned out to be first. In his exercise, Johnson had built, as Kennedy had wanted, a strong justification for a presidential initiative to undertake Project Apollo, but he had also moved further, toward a greater consensus for the objective among key government and business leaders. NASA's leaders were enthusiastic about the course Johnson recommended. James E. Webb (1906–1992), NASA chief administrator from February 1961 through October 1968, with powerful connections to Congress, made sure that NASA was supported fully and on the long run, recommending that the national space plan included the objective of manned lunar exploration within the given decade (Lambright, 1995; Bizony, 2007). As is well known, on May 25, 1961 Kennedy hold a second State of the Union address asking for the most open-ended commitment ever made in order to land an American on the Moon (Saari et al., 2005; Beschloss, 1997). Was it as enthusiastically though, as Launius (2004) writes, that Congress appropriated funding for Apollo?

In 1963, dissenting views grew louder, taxpayers complaining about the costs, scientists resenting the slighting of more important project (Etzioni, 1964). *The New York Times,* reflecting public opinion, proclaimed the lunar program to be in crisis (The New York Times, July 13, 1963). In the fall of 1963 at the United Nations, Kennedy then made a serious case that the Unites States and the Soviet Union should explore the Moon together (see his Memorandum for NASA Administration of November 12, 1963, in Launius, 2004).

Kennedy's commitment to space nevertheless captured the American imagination and attracted overwhelming support. No high official at the time seemed deeply concerned about either the difficulties or the expenses. It was essentially due to Johnson's effectiveness – first as vice president, and after November 1963 as president – in building a nation's consensus for a space program credit that this commitment developed. Observing some of the most powerful people endorse the idea of a space program made it easier for the rest to follow (Dallek, 1997). Johnson increased NASA's spending by 150 million to over 4 billion USD, not so much anymore in order to win the race for space, but to develop the technology and the science of space in order to be present. Johnson also believed that the Apollo mission made excellent economic and political sense. Land a man on the Moon would not only reaffirm America's superiority, so was Johnson's saying, but would also spur both immediate and long-term economic growth and gain the administration considerable political credit with the public. And in 1963, it was Johnson also who amplified the appropriation for the space program, by pointing out that the space program as expensive as it was, could be "justified as a solid investment which will give ample returns in security, prestige, knowledge, and material benefits" (L. B. Johnson cited in Dallek, 1997: 73). It was of little use that economists argued against it (Solo, 1962; Smith, 1963). Polls carried out in 1964 onward showed that he had made his point.

Likewise argued NASA's chief Administrator Webb. He kept close track of how NASA affected the nation's economy and took every opportunity to apprise Johnson of this gain. In a 1965 report to the president, Webb pointed out that, in the previous year, 94 percent of NASA's investment had gone to 20,000 private US industrial companies, 331 million had been spent in 120 cities in 22 states with high unemployment rates, and as many as 750,000 people worked directly or indirectly on NASA related business. What's more, he delineates that new materials, complex electronics, mechanicals and chemical systems were developed, bringing with it revolutionary change in the way of making and testing things, not only for space systems, but for innumerable other non space services, processes and materials (Dallek, 1997). Webb was very enthusiastic, and eager in accomplishing what Johnson had declared as a vision of a "Great Society": "I know of no area", he told the president, "where the inspirational thrust toward doing everything required of a great society can be better provided on a proven base of competence, and with so many practical additional benefits to be derived, than through the space program." (J. E. Webb cited in Dallek, 1997: 77–78).

Johnson's inclination to be generous with NASA and provide for a modest amount of post-Apollo spending could withstand neither a disastrous fire in an Apollo command module in January 1967 nor a growing US budget deficit spurred by the fighting in Vietnam. The destruction of *Apollo 1* and the killing of the three astronauts undermined further national confidence in NASA, which was for some time accused of carelessness in trying to move project Apollo forward too quickly (Etzioni, 1964; Dallek, 1997). Senate hearings raised questions about a large number of defects in the spacecraft. Webb's decision for a straightforward policy in responding to the failings that produced the fire restored a measure of confidence in the agency and prompted the Senate committee to recommend that NASA continue to move on the Apollo program (Dallek, 1997; Bizony, 2006). Nevertheless – and/or as a consequence? – NASA's budget declined with fiscal year 1967 and did not even turn around with the successful lunar landing. Webb brought forward the cold war rhetoric to press for an increase of NASA budget, however in vein; the rhetoric had lost its power. In fact, as Hoff (1997) has pointed out, the July and November Moon landings probably diminished rather than contributed to furthering public interest, and hence sustained the disillusionment felt by many Americans about Project Apollo. While technologically innovative and visually exciting, the program left much to be desired from most other vantage points.

Richard Nixon (1913–1994), who took the presidency in 1969 after Johnson, first showed an initial outburst of enthusiasm by declaring the flight of *Apollo 11* as the representation of the "most significant week in the history of Earth since the creation" (R. Nixon cited in Launius, 2007: 36). Nixon eventually moved space technology away from being a political and military weapon in the cold war. In addition to severe economic problems that Nixon had inherited from the massive spending programs in Vietnam and the Great Society, it seems that Nixon just never exhibited the personal enthusiasm for an expansive commitment to the space program that Johnson and Kennedy had shown. In fact, the post Apollo era had already been restricted by Johnson, who had only agreed upon funding for the Apollo program, and never sincerely considered a post-Apollo era (Dallek, 1997).

Subsequently, Nixon did not perceive the space program to be in crisis as a result of lower funding, unlike other domestic issues. Apparently, he revealed no particular interest in the subject at all. He had mentioned the space program more often during his 1960 presidential campaign than he did in the 1968 campaign. Although he spoke of the *Apollo 11* accomplishment as the "most exiting event of the first year of my presidency" (R. M. Nixon cited in Hoff, 1997: 97), his presidential papers document clearly that his personal interest was more in the diplomacy of space.[3] All these factors indicate his basic agreement with a moderate level of spending in this arena and his preoccupation with other issues. Modest presidential enthusiasm resulted in little attention overall for the space program.

This historical tale illustrates the now well-documented observation that the spread of an idea depends highly on the enthusiasm of a few key individuals with the necessary power and influence to pursue it, ready to take risks to push the plan forward and adapt it in innovative ways to the context (Gladwell, 2002). Kennedy and Johnson had to face serious political and economic issues too. Yet they were enough attached to the idea to send a man to the Moon that they stuck with the idea, and pursued it sufficiently vigorously to involve other people to broaden the support.

### 3.2 Funding

The human spaceflight program was undertaken by NASA from 1961 through 1975, Apollo spearheading NASA's overall program during the sixties.[4] Although the lunar landing generally overshadowed other important activities, critics of the agency often saw the near term goal as an end in itself, while the program stimulated phenomenal progress in aerospace technology. Building upon the pioneering achievements of Mercury and Gemini, Apollo produced dramatic advances in launch vehicles, spacecraft and operational techniques.

In 1961, Kennedy managed to sign in as much as eighteen contractors. North American Aviation Inc., a major contractor being responsible for the manufacturing of three spacecrafts, signed a contract of 900 Million USD for the period from August 1961 through November 1962 (Johnson, 2007). Others did not get as much, however the contractors all together obtained 1.5 billion USD over the same time period, ready to be incorporated into the Apollo program (Ertel and Morse, 1969). As 1962 drew to a close, NASA had designed the most costly and challenging engineering project in civil times. In February 1963, NASA signed a contract with Boeing for about 419 Million Dollars to build ten boosters.[5] It has to be emphasized though that most contractors kept other businesses on track in order not to be too dependent on the space industry (Diamond, 1964).

An overview of the funding during the fiscal years 1961–1969 highlights the importance of the Apollo project. (Note that it is quite difficult to obtain budget figures and facts on funding. Whereas the publications and research reports on technical details of the Apollo project constitute an enormous body of literature, only little has been published on the financial aspects so far.) The Apollo share of the total NASA Budget increased over the years between 1962 and 1970 from 10% (1962) to 70% in 1967 (all figures are drawn from Ertel and Morse, 1969–1978; see Table 1). Financial support wasn't a given thing, though, but had to be negotiated on a yearly basis: "People frequently refer to our program to reach the moon during the 1960s as a national commitment," Lyndon Johnson once wrote, "It was not. There was no commitment on succeeding Congresses to supply funds on a continuing basis. The program had to be justified and money appropriated year after year. This support was not always easy to obtain." (Johnson, 1971: 283).

| FY | NASA USD 1960s value | NASA USD 2006 value | Apollo USD 1960s value | Apollo USD 2006 value | GDP USD 1960s value |
|---|---|---|---|---|---|
| **1961** | 964 | 5'280 | – | – | 544'700 |
| **1962** | 1'672 | 9'034 | 160 | 865 | 585'600 |
| **1963** | 3'674 | 19'645 | 617 | 3'300 | 617'700 |
| **1964** | 3'975 | 20'934 | 2'273 | 11'970 | 663'600 |
| **1965** | 4'271 | 22'088 | 2'615 | 13'523 | 719'100 |
| **1966** | 4'512 | 22'688 | 2'967 | 14'922 | 787'800 |
| **1967** | 4'175 | 20'366 | 2'916 | 14'225 | 832'600 |
| **1968** | 3'971 | 18'576 | 2'556 | 11'958 | 910'000 |
| **1969** | 3'194 | 14'234 | 2'025 | 9'026 | 984'600 |
| **1970** | 3'114 | 13'181 | 1'686 | 7'137 | 1'038'500 |
| **1971** | 2'555 | 10'300 | 914 | 3'683 | 1'127'100 |
| **1972** | 2'508 | 9'689 | 601 | 2'323 | 1'238'300 |

Table 1: Authors' compilation, based on the Final Budget Appropriation for NASA and the Apollo Project (in Million Dollars) taken from Ertel and Morse, 1969–1978; GDP data (in Million Dollars) taken from http://www.bea.gov/index.htm [URL April 30, 2008].

Yet, the financial funding of NASA increased yearly between 1961 and 1966, at which time it reached its peak. Correspondingly, the share of the Apollo program also increased with each year. In 1963, the final fiscal budget for Apollo was still 0.62 billion USD or 17% of the entire NASA budget. In 1964, it increased to 2.27 billion USD or 57% of the overall NASA budget, even though the aim of putting men on the Moon was still far beyond reach. The public complaining about the costs in 1963 obviously did not put an end to the support by Congress; it merely led to a momentary halt of the budget increase. The two years 1963 through 1964 saw the essential completion of the putting together of the Apollo government-industry terms, a substantial maturing of the design, the verification of many essential design features by tests, and streamlining of the flight program. In 1965, the total sum of 4.27 (and not 5.2 billion USD as is given in numerous summaries in the literature) went to NASA, and thereof 2.61 billion USD to the Apollo program. The funding reached its peak in 1966 with an overall NASA budget of 4.51 billion USD, whereof 2.97 billion USD or 67% went to the Apollo program.

An accident with *Apollo 1* on January 27, 1967 (a fire in the crew module, bringing three astronauts to death) was a severe blow to NASA. The accident forced a temporary halt to the program and the agency's safety procedures underwent extensive review, the budget for Apollo slightly decreasing to 2.60 billion USD in 1968. Nevertheless, the first manned flight was *Apollo 7* in October 1968 and in 1969 two men were put on the Moon. However, in the early 1970s, after a few more manned flights, interest in furthering Moon exploration waned. After the initial rise of efforts embodied in the Apollo program, space exploration reached an equilibrium, accompanied by drastic budget shortening; the fiscal budget in 1971 was 0.91 billion USD for Apollo or 36% of the overall NASA budget, and in 1973 it was down to a minuscule 3% fraction only. The last Apollo mission landing astronauts on the Moon was in 1972, when *Apollo 17* concluded the Apollo mission. After that the United States did not undertake any other Moon flights (Nagel and Hermsen, 2005). Further on, NASA concentrated its efforts on the launch of reusable spacecrafts, i.e. the space shuttle program, on Space Telescope projects and on the International Space Station. As of 2008, there has not been any further human spaceflight beyond low Earth orbit since the last mission in the Apollo program. Concomitantly, the scale of funding reached during the Apollo program has never been seen again (Orloff and Harland, 2006).

## 4. Human and social dimensions

As interesting as it would be, it is out of the scope of this paper to recount the whole story, which can anyway be found in many sources cited in our bibliography. For the purpose of testing the pro-bubble hypothesis, we are interested in identifying situations in which people exhibited a propensity to take large risks associated with an initiative or for the development of a project. The Apollo program was a culmination of the work of hundreds of thousands of people. Most of them remain unidentified. We thus have to focus only on a few exceptional characters, narrating parts of their biographies that put in perspective their enthusiasm for the space project. We will take into account the high risks that the protagonists were ready to accept. We claim that such projects were carried out with a rising (collective) enthusiasm, and willingness for undertaking such major risks both financially and personally. This will lead us to a second subsection which describes the dynamics of the public support for the Apollo program.

### 4.1 The Protagonists

In the early years of the space program, expectations were high, and so was the eagerness to be the first in space as well as being in space at all. Not only politicians, but the scientists and engineers themselves were more than enthusiastic to invest everything to succeed.

One of the most important figures among the early NASA employees was the German native Wernher von Braun (1912–1977) (Cadbury, 2005; Saari et al., 2005; Becker, 2007; http://history.nasa.gov/sputnik/braun.html [URL: April 30, 2008]). As a youth, he became enamored with the possibilities of space exploration by reading the science fiction novels of Jules Verne and of H. G. Wells, and the science writings of Hermann Oberth (*Die Rakete zu den Planetenräumen [By Rocket to Space],* 1923). From his teenage years, von Braun had held a keen interest in space flight, becoming involved in the German rocket society, *Verein für Raumschifffahrt*, as early as 1929. As a mean to further his desire to build large and capable rockets, in 1932 he went to work for the German army to develop ballistic missiles. Von Braun is well known as the leader of what has been called the "rocket team", which developed the V-2 ballistic missile for the Nazis during World War II. The V-2s were manufactured at a forced labor factory called *Mittelwerk*. Scholars are still reassessing von Braun's role in these controversial activities.

The invention of the V-2 rocket by von Braun's rocket team, operating in a secret laboratory at Peenemünde on the Baltic coast, was the immediate antecedent of those used in space exploration programs in the United States.

By the beginning of 1945, von Braun began planning for the postwar era. Before the Allied capture of the V-2 rocket complex, von Braun engineered the surrender of 500 of his top rocket scientists, along with plans and test vehicles, to the Americans. For fifteen years after World War II, von Braun would work with the United States army in the development of ballistic missiles.

In 1960, his rocket development center transferred from the army to the newly established NASA and received a mandate to build the giant Saturn rockets. Accordingly, von Braun became director of NASA's Marshall Space Flight Center and the chief architect of the Saturn V launch vehicle, the superbooster that would propel Americans to the Moon. That's where he wanted to be since 1945: "We've got mighty little time to lose, for we know that the Soviets are thinking along the same lines. If we do not wish to see the control of space wrested from us, it's time, and high time, we acted." (von Braun, 1955, cited in Cadbury, 2005: 147). In 1959, official approval was given for the Saturn project, which aimed to develop a booster rocket with enough power to reach the Moon (Cadbury, 2005).

Von Braun's fame, though, vanished after 1965, when there was continued opposition to the mounting cost of the Apollo program, the engineer becoming a target for popular criticism, due to his money consuming efforts as well as his past as a Nazi-Member, something he had drawn the curtain over (Cadbury, 2005).

Enthusiasm was soaring with other scientists and engineers too. In the 1950's already, in the Apollo forerunners projects, people were eager to test their engineering developments. Since tests with small animals such as hamsters and mice were not sufficient to answer all questions, self-experimenting was thought of as a matter of course. In order to test the impact of weightlessness on the human body (something we know nowadays to be low-risk, however not so at the time) (Stuster, 1996), self-experiments were a daily common endeavor. For instance, Herbert D. Stallings estimated that by April 1958, he had flown more than 4000 zero-g trajectories and compiled about 37 hours of weightless and subgravity flight in order to prove its harmlessness (http://history.nasa.gov/SP-4201/ch2-3.htm [URL April 30, 2008]).

John P. Stapp (1910–1999) from the Aeromedical Field Laboratory was a pioneer in studying the effects of acceleration and deceleration forces on humans. When he began his research in 1947, the common wisdom was that a man would suffer fatal internal injuries at around 18 g. Stapp shattered this barrier in the process of his progressive work, experiencing more "peak" g-forces than any other human.

Stapp suffered from repeated and various injuries including broken limbs, ribs, detached retinas, and miscellaneous traumas which eventually resulted in lifelong lingering vision problems caused by permanently burst blood vessels in his eyes. In 1953, he conducted a sequence of self-testing where he was seated on a rocket sled without and with helmet. He tested speeds of several hundred km/h. When decelerating he endured a force of 40g (see figure 1) (Becker, 2007).

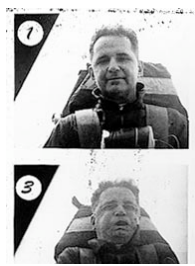

Figure 1 (taken from Becker, 2007: 28).

Putting one's own life at stake for space science was also beyond discussion for Virgil Ivan Grissom (1926–1967). Grissom was well aware of the risks involved in the spacecrafts he was boarding, and he always felt that they were a calculated risk. Grissom was an Air Force pilot and a NASA astronaut (http://www.jsc.nasa.gov/Bios/htmlbios/grissom-vi.html [URL: April 30, 2008]). He was the second American to fly in space. He had entered the Air Force Institute of Technology to study aeronautical engineering, and was pilot of the second American suborbital spaceflight (Mercury Redstone 4). Following the splashdown of the craft, the hatch, which had explosive bolts, blew off prematurely, letting water into the capsule and into Grissom's suit. Grissom nearly drowned but was rescued, while the spacecraft sank in deep water (Becker, 2007). In early 1964, Grissom was designated command pilot for the first manned Project Gemini mission, making him the first astronaut to return to space as well as the first person to fly twice above the accepted boundary of space in a capsule-based spacecraft.

When Grissom and fellow astronaut John Young walked out in March 1965 to Gemini 3, the risks were much larger than in previous flights. This was a new launch vehicle, a new larger spacecraft and the first flight for humans in "space." In previous rockets, there was an escape tower. In the new Gemini 3, the crew escape system had been reduced, diminishing its capability. The new flight was a big, risky step, perceived as such by the astronauts, while the public was largely unaware of these new risks. A few days before the launch of the Gemini 3 aircraft, Grissom alarmed NASA administrator James Webb by saying: "If we die, we want people to accept it. We're in a risky business, and we hope that if anything happens to us, it will not delay the program. The conquest of space is worth the risk of life." (V. I. Grissom, March 1965, cited in Barbour *and* Verecken, 1969: 125). His career with NASA confirms his statement as a testimony of the foundation of his commitment. Grissom was made commander of a Saturn Launch vehicle, intended to be the first manned Apollo flight. On January 27, 1967, when running through a series of tests for the *Apollo 1* mission in a simulated flight environment, he was killed together with two fellow astronauts (Ertel and Morse, 1978). The deaths of the three astronauts were attributed to a wide range of lethal design hazards in the early Apollo command module such as its highly pressurized 100% oxygen atmosphere during the test, many wiring and plumbing flaws, flammable materials in the cockpit, a hatch which might not open at all in an emergency and even hazard from the flight suits worn by the astronauts. An internal NASA enquiry resulted in the spacecraft problems being fixed and successful resumption of the Apollo program (http://www.jsc.nasa.gov/Bios/htmlbios/grissom-vi.html [URL: April 30, 2008]).

Not least committed to space flight was John Herschel Glenn (1921–), in his youth as well as in later years. He was the third American to fly in space and the first American to orbit the Earth in February 1962 http://www11.jsc.nasa.gov/Bios/htmlbios/glenn-j.html [URL: April 30, 2008]). After retiring from the Marine Corps on January 1, 1965, he was a business executive from 1965 until his election to the United States Senate in November 1974 (retired 1999). Decades later, at the age of 77, Glenn lifted off for a second space flight on October 29, 1998, on a Space Shuttle. His intention was to study the effects of space flight on the elderly people (Becker, 2007). He became the oldest person ever to go into space. Whereas some criticized Glenn's participation in the nine-day mission as a junket for a politician, others noted that Glenn's flight offered valuable research on weightlessness and other aspects of space flight on the same person at two points in life.

Altogether, being part of space technology was much more than participating in the space race. Many parties involved expressed their decision to take part as a result of their desire and passion. Building Apollo was more than just the yearning to beat the Russians. It was the fulfillment of a dream, be it a boy's dream, a dream fueled by the heaps of Science fiction narratives (Dick and Cowing, 2004) or even the vision to build a better society in space (Launius, 2006). And it did not free the impassionate participants until the dream of the lunar landing was fulfilled. As *Los Angeles Times* reporter Rudy Abramson has put it in 1969 – just three days before the *Apollo 11* launch –: "The United States this week will commit its national pride, eight years of work and $24 billion of its fortune to showing the world it can still fulfill a dream." (R. Abramson, Los Angeles Times, July 13, 1969, cited in Collins, 1974: 315).

After all, taking risks was not seen as putting one's life in danger, but getting the chance to explore things, to experience space exploration. "My desire [to explore] outweighed the risk I perceived, a risk greater than I probably realized at the time", said Michael Foale (cited in Dick and Cowing, 2004: 262), an astronaut with NASA in the 1980s. And since the astronauts felt that risk was always with them, their interests were focused on the rewards coming along with exploration rather than on risk – corresponding to an upward biased perception of the *reward/risk ratio*, typical of the bubble spirit we wish to document. Some did it because they felt they wanted to contribute to innovative science (Dick and Cowing, 2004), others just for the sheer curiosity. This was the case for Shannon Lucid, Space Shuttle astronaut in the late 1970s: "[…] why I really wanted to go was because I was curious. I was very curious to find out what it would be like to live and work in space for a long period of time, I was curious to see how the body would adapt. And I really wanted to experience that." (S. Lucid cited in ibid.: 161). Risk was something not to be ignored, but to be handled: "I took the perceived risk, or took what I felt was a risk, and changed it into something I thought I could handle. I changed it into a procedure. I forced the system into accommodating what I could handle. And so that's how I handled those risks." (ibid.: 163). Risk was not to be ignored, but to be aware of, and to be reduced: "After the risk is past, crew members, family, space managers, all of us are relieved, and we celebrate how we have cheated death once more. It shows in our faces that the risk of spaceflight and space exploration is always present, and we must always be honest about it, explain it, and do our utmost to reduce it, without hiding it. […] No, you talk about it." (M. Foale cited in ibid.: 266; see also Lovell, ibid.). These quotes show not much of a difference between the earlier protagonists, and their followers. The risk had to be taken in any case, for the sheer sake of space exploration. This was common wisdom at the time: "The greatest risk is not to explore at all." (ibid.: passim).

### 4.2 The Public Response

Launches from Cape Canaveral inevitably drew hundreds of thousands of excited spectators; public support of the space program in its early stage was enthusiastic. It was the time when Congress decided that NASA, and not NASA *and* the Air Force, would put a man on the Moon. During the negotiations, John Glenn and Scott Carpenter had orbited the Earth, and the American public was cheering for its new space agency. In 1962, the high volume of sales of space toys and kits in department stores after each launch was proof of the public attraction to anything related to space (Diamond, 1964). As early as 1963, however, criticism was beginning to grow in the press. Whereas in 1962, the editors used superlatives when talking of space exploration (e.g. Fortune, June 1962), in 1963, enthusiasm was somewhat low (Smith, 1963; Diamond, 1964). The lavish amount of money being poured into NASA was being questioned, should it go up constantly (Fortune, June 1962; Smith, 1963). A number of writers criticized the program as a cynical mix of public relations and profit-seeking, a massive drain of tax funds away from serious domestic ills of the decade, a technological high card in international tensions during the cold war era (Fortune, June 1962). The last orbital flight had been in May 1963 and it looked very ordinary compared to the Soviet double act. A series of congressional hearing questioned the value of a lunar program (Cadbury, 2005). Among scientists, the initial enchantment had faded before the mounting costs and they feared the heavy drain on other fields of scientific endeavor. Furthermore, less and less was heard of the military urgency of exploring the Moon, even the Air Force decided its interests would lie more in "innerspace" capability. Payoffs of the Program seemed to vanish (Smith, 1963). As a consequence, in July 1963, *The New York Times* carried a headline proclaiming *Lunar Program In Crisis*, echoing popular sentiments (The New York Times, July 13, 1963). What has happened? The impetus for this declaration was not only managerial problems within NASA but foremost a study concluding that the odds of reaching the Moon by the end of the decade were only about one in ten and that a lunar landing could not be attempted "with acceptable risk" until late 1971. Others criticized the Apollo program for its timetable, for being developed in the context of a race rather than as an undertaking following a reasonable pace. The timetable derived from the race called for bringing it to a culmination in 1967 or 1968. However, this speedup increased the costs tremendously (Smith, 1963; Etzioni, 1964). What's more, 1964 was the year when astronauts expressed doubts on the skills of scientists and engineers (Collins, 1974), maybe as an answer to the reservations expressed by the scientists for potential gains coming from manned space exploration.

All the same, the pendulum again swung towards the space program. After this temporary drop of enthusiasm in 1963, polls in the spring and fall of 1964 showed 64–69 percent of the public were favorably disposed to landing an American on the Moon, with 78 percent saying the Apollo program should be maintained at its current pace or speed up. Arguments on both economic and technological levels paid their tributes.

Polls performed in summer 1965 showed a new decrease to a certain extent, since a third of the nation now favored cutting the space budget, while only 16 percent wanted to increase it. Over the next three and a half years, support for cutting space spending went up to 40 percent, with those preferring an increase dropping to 14 percent. At the end of 1967, *The New York Times* reported that a poll conducted in six American cities showed that five other public issues held priority over efforts on outer space. Residents of these cities preferred tackling air and water pollution, poverty and others before spending additional federal funds to space exploration. The following year *Newsweek* echoed the *Time's* findings, stating that the United States space program was in decline.

At the same time, Congress was strongly disposed to reduce NASA's budget. A White House survey of congressional leaders at the end of 1966 revealed pronounced sentiment for keeping Apollo on track but, simultaneously, for cutting NASA spending by skimping on post-Apollo outlays. Space age "seemed truly to have arrived," stated astronaut Collins, that year (Collins, 1974: 255). In this context, in January 1967, Johnson requested a 5 billion USD NASA budget for Fiscal Year 1968 (Dallek, 1997). It has already been stated, however, that 1966 was the peak of funding for both NASA and the Apollo program, and Johnson did not succeed as he might have wished.

The efforts of 1966/67, however, paid their tributes: On Christmas morning of 1968, the Americans were greeted by newspapers, radio and television reports with the momentous news that the crew of *Apollo 8* was on its way back to Earth after becoming the first human beings to orbit the Moon. For the United States space program, this endeavor represented a major step toward achieving the national goal, and gaining support with it. The news that three astronauts had flown around the Moon sparked feelings of national pride. This event stood in stark contrast to the previous events of 1968 (the Vietnam war still going on, the assassinations of M. L. King and R. Kennedy, among others). The nation was becoming increasingly divided over the issues escalating during this year. *Apollo 8* provided an uplifting end to all these negative events. One of the telegrams the astronauts received from the public summarizes well the general spirit associated with *Apollo 8* mission: "You [i.e. the astronauts] saved 1968." (Chaikin, 2007: 54–55).

In particular, this event created a general new awareness of the Earth, as representing "a grand oasis in the big vastness of space", as astronaut Lovell expressed it, overheard by millions of people during a television broadcast by the astronauts from lunar orbit. This made *Apollo 8* different from previous explorations. Many observers have noted the coincidence between *Apollo 8* and increase in environmental activism (Bizony, 2006; Chaikin, 2007).

The historical impact of *Apollo 8* was only equaled when *Apollo 11* made it to the Moon on July 19, 1969. An estimated 600 million people – one fifth of the world's population – witnessed it on television and radio. Some observers designated the day as a turning point in history. Scientific writer Robert Heinlein who had penned the story for the 1950 film *Destination Moon* named it the "[…] greatest event in all the history of the human race up to this time." (R. Heinlein cited in Chaikin, 2007: 55).

After *Apollo 11* had landed on the Moon, lunar scientists as well as astronauts became highly optimistic about the outcome of scientific research associated with orbiting flights and exploration of the Moon. Astronauts had demonstrated that men were able to function as explorers in the lunar environment (Michael Collins, as an example). They were viewed by the advocates for manned space flight as ample justification for the enormous investment they required. Hopes of the scientists for resolving major questions about the origin and evolution of the Moon reached a peak of optimism at the beginning of 1970.

### 4.3 The aftermath

The international chariot race, however, ended when the Eagle landed (Lewis, 1974). With astonishing rapidity, the raison d'être of the Apollo program had undergone a metamorphosis. Overnight, it transformed into a scientific undertaking for the highest intellectual purpose. By the spring of 1970, it was obvious that the intellectual rationale for Apollo could not justify the full program in the absence of enthusiastic public support, and that was waning. In November 1969, *Apollo 12* astronauts achieved a second lunar landing and made two Moonwalks. Once again, there were live pictures, but the news coverage was showing signs of apathy. One Tennessee resident was even quoted as naming the event an "old hat" (Chaikin, 2007: 58). The second landing on the Moon lacked of enthusiasm, the public progressively more disenchanted with the space program. The voice of Apollo's critics – always there but never very loud (see Diamond, 1964; Etzioni, 1964; Bilstein, 1989) – thus began to swell in volume. It became an amusement to ridicule the program in intellectual and political circles; support began to wither (Lewis, 1974). The national polls in the summer of 1969 found that 53 percent of the country was opposed to a manned mission to Mars. And a poll taken in 1973 showed that only foreign aid had less support than space exploration (McQuaid, 2007). Failures of the *Apollo 13* mission added fuel to the fires of criticisms. Also within NASA, the administrators sensed that subsequent missions could not afford any other failure.

There were several factors that exacerbated the decline in interest. The society was changing. Chaikin (2007) stresses the fact that unlike science fiction writers and their readers, most Americans had little familiarity with space technology and although TV commentators struggled to convey the nuts and bolts of space exploration, arcane concepts like space rendezvous were, literally and figuratively, over the viewer's heads. In addition, NASA tended to emphasize the technical elements of the program rather than the human experiences that would have been easier to understand. Lunar science became increasingly the focus of astronauts as well as mission planners. For non-scientists, geologic talk about the Moon and the origin of Earth wasn't easy to follow. The cultural divide between scientists and the rest seemed to grow (Chaikin, 2007). From a practical standpoint, only a handful of Earth scientists was benefiting from Apollo to pursue research of questionable social relevance at enormous expense, to the deprivation not only of more immediate and pressing social needs, but also of other "relevant" scientific goals. This attitude spread through many sectors of American society that had supported space research enthusiastically a few years earlier, notably academia. Liberals equated the funding of technological development and fundamental scientific inquiry (except in the field of medical research) as being indifferent to social welfare. The same attitude was adopted by the military establishment and its committees in Congress. A substantial portion of the radio astronomy program in the United States that the Department of Defense had supported for years promptly collapsed. The Apollo program, which was born during the Democratic Administration of John F. Kennedy in 1960, had virtually been disowned by the liberal wing of the Democratic Party by 1970 (Lewis, 1974). And President Nixon's intentions – as we have seen above – were far from helping to change this course. Even though the *Apollo 17* astronauts felt that the lunar expeditions were instilling confidence among the American people (Dick and Cowing, 2004), after *Apollo 17*, the United States did not undertake any more lunar flights. Thomas Paine, successor of James Webb as NASA Administrator, had to cancel missions 18, 19, and 20, in order to save money, and assuage the new President (Bizony, 2006).

In the short term, Project Apollo was an American triumph. In the long term, the costs, close to 25 billion USD overall in 1960s dollars, were large and might have made a difference in other programs or helped avoid the inflation that fueled dissatisfaction during the Vietnam War (Chaikin, 2007). Furthermore, we now know that the reason for the Soviet Union loss of the cold war was that it could not compete with Western financial and corporate power. From the beginning, the Soviets were behind in almost every kind of technologically complicated armament. Actually, the Soviet Union took its position in space out of weakness. It developed its space program based on less advanced technology (but more robust as shown by the exceptional robustness of its MIR space stations and the use of its old technology launchers by the United States and many other nations after the end of the cold war). On the American side, the lack of transparency of the USSR produced a series of efforts to find out what the Soviets were doing. The Sputnik I launch led directly to that of the first reconnaissance satellite by the United States early in 1961, and in turn inspired the Soviet to other space endeavors (Ferrell, 1997). The moves and countermoves did not all fit together neatly, but the Soviet accomplishments brought the Americans into a full-scale, open race for the Moon. Although the Moon program contributed a great deal to the United States, one could argue that the tens of billions of dollars spent in the 1960s on what Kennedy essentially thought of as World propaganda could have been otherwise devoted to the United States domestic economy or even defense, and that might have convinced the Soviets more quickly of the fruitlessness of the tragic conflict with the United States.

## 5. Conclusions

Today, more than three decades after the program ended, Apollo remains a unique event in the history of space exploration. It was one of the most exceptional and costly projects ever undertaken by the United States, and thus constitutes an excellent example of how bubbles function from within. In the context of bubbles associated with innovation ventures and the creation of new technology, the Apollo Project demonstrated the large risks that have been undertaken individually, politically and financially, leading to a collective (individual, public and political) over-enthusiasm, which played a very significant role in the development and completion of the process as such. The qualifier "over" emphasizes that the enthusiasm did not out-live by much the first Moon landing, and a general positive sentiment in favor of the Moon exploration started to fade shortly after the first step on the Moon. The evidence gathered here supports the view that the Apollo program was a genuine bubble, in the sense of our general definition expressed in section 1, with little long-term fundamental support either from society or from a technical or scientific viewpoint. As expected from our hypothesis on bubbles, it led to innumerable technological innovations, and scientific advances, but many of them at a cost documented to be disproportionate compared with the returns. These returns may turn out to be positive on the long run as many of its fruits remain to be fully appreciated and exploited.

With the first landing on the Moon in 1969, it was the general belief at the time that thirty years later at the transition to the third Millennium, mankind would have established permanent stations on the Moon and on Mars, with space travel expected to become almost routine and open to commercial exploitation for the public. With the hindsight of 2008, it is easy to dismiss this view. Here, we stress instead that this view exemplifies the bubble spirit which is typical rather than exceptional. Enthusiasm was always present to push for the endeavor, risk always a topic but never an issue. Major risks have been accepted by individuals, first of all by some of the pioneers (engineers, astronauts). They did not shy away from taking all possible types of risks, even at the costs of their own life or health. Such high risk levels were accepted for "real reasons," in the sense summarized buy today's NASA administrator Michael Griffin in January 2007. Real reasons are intuitive and compelling, but they are not logical, they are not the standard acceptable logical reasons based on solid rational cost/benefit analysis (Griffin, 2007). Real reasons are the opposite; they include curiosity, quest for the fulfillment of dreams, competitiveness (because people want to leave a legacy), and challenge, for the sheer reason that it's there. Even during Apollo's last years, the launch team's esprit de corps seemed to be of central concern. It could in fact be seen as a personal commitment of each member of the launch team (Benson and Faherty, 1978).

Enthusiasm shown by the public and by the financially responsible entity, the Congress, was on the other hand somewhat more complex. Commitment by the public was swaying. It did not always stand with NASA in equal measure, thus was not always agreeing on taking the risks that went with investing in such a major endeavor. In the early stage of Apollo, support was undivided, and the first concerns arose in 1963. In 1964 though, enthusiasm and support with the program were back (69% positive). Arguments on both economical and technological levels obviously paid their tributes, most likely due to the effort of Johnson and Webb after 1963. The summer 1965 then showed a certain decrease, since a third of the nation now favored cutting the space budget (66% positive). Over the next three and a half years, support for the space program went down to 60 percent (60% positive), and it only increased for a short time during the lunar landing in mid 1969.

At the same time, Congress acted somewhat differently. Approved funding was rising until its peak in 1966. The "upheaval" by the public in 1963 had no direct manifestation in Congress, on the contrary, at the time it was strongly willing to increase NASA's budget. After 1966 though, support slowly but steadily decreased until it was cut off almost entirely under Nixon. Since Nixon was not interested in pursuing the "Apollo idea", it terminated in 1972. This ending though did not come as surprising as it seemed at first sight, since its decline loomed on the horizon some years before, after 1965 from the public's, after 1967 from Congress' side.

We have called the Apollo program one of the most exceptional and costly project ever undertaken by the United States at peacetime. However, just as Apollo had come out of nowhere, and held center stage for a decade, it vanished from the public consciousness, as if it had never happened. It thus was an exceptional niche, not having been revisited to any significant degree ever since.

In fine, from the point of view of the pro-bubble hypothesis, a key question concerns the identification of the main outcomes of the program. In a nutshell, what were and what are the returns resulting from the Apollo bubble? As we all know, several successive programs followed Apollo. After a six-year hiatus in America's human spaceflight missions, in April 1981, NASA achieved the first flight of its reusable Space Shuttle. Buoyed by NASA's promise that the Shuttle would make spaceflight routine, many people responded once more with high enthusiasm. And in the first several years of the Shuttle missions, there was plenty of action to excite space buffs: they could witness spacewalking Shuttle astronauts repairing satellites and flying through the void with self-propelled maneuvering units. The fact though that the Shuttle never ventured beyond low-Earth orbit was lost on those Americans for whom spaceflight had become synonymous with going to the Moon. As for the 21st century, NASA is now slowly gearing up to resume exploration to the Moon (Harland, 2008). In January 2004, President G. W. Bush announced a new vision for Space exploration, including the return to the Moon, and eventually going to Mars (Dick and Launius, 2006; Johnson, 2007). And if all goes according to plan, astronauts will be back on the Moon no later than 2020 (Chaikin, 2007).

Space exploration is thus a robust manifestation of a venture where technological innovations rely on enthusiasm leading to risk taking on a collective level. Further efforts should now include the questions of how such major projects can increase social welfare. Is it possible that nascent, emerging industries need "animal spirits" and over-investment for innovation? Or is it more that bubbles serve mainly to change the wealth dynamics of society, and through this mechanism affect the investment process? Much more work is needed in this concern to clarify these questions.

- http://history.nasa.gov/SP-4201/ch2-3.htm [URL April 30, 2008].
- http://history.nasa.gov/sputnik/braun.html [URL: April 30, 2008].
- http://www.jsc.nasa.gov/Bios/htmlbios/grissom-vi.html [URL: April 30, 2008].
- http://www.nasa.gov/topics/history/index.html [URL: April 30, 2008].
- http://www11.jsc.nasa.gov/Bios/htmlbios/glenn-j.html [URL: April 30, 2008].

---

[1] The literature on the Apollo project is huge, first of all since NASA itself has produced (and still is producing) a large body of it, in the form of the NASA History Series; for an overview see Garber (1997); others are Logsdon (1970); Benson and Faherty (1978); Brooks, Grimwood and Swenson (1979); Murray and Blay Cox (1989); Bilstein (1989, 1994, 2003); Nagel and Hermsen (2005); Orloff and Harland (2006). The majority of these books are focusing on technical issues; furthermore there exists some good compilations with a critical mindset, and on political, managerial and societal issues, see Diamond (1964); Etzioni (1964); Logsdon (1997); Launius and McCurdy ed. (1997); Dick and Cowing (2004); Cadbury (2005); Saari, Baker and Hermsen (2005); Saari et al. (2005); Dick and Launius ed. (2007); Harland (2008).

[2] In comparison, the *Manhatten Project*, established during World War II for the development and construction of the Atomic Bomb, required 2 billion USD.

[3] After the 1970's, international competition lessened, and international cooperation between the United States, Canada, and the nations of Western Europe stepped forward. For a comprehensive summary and evaluation of the space program see Ferrell, 1997: 172–204.

[4] NASA was established in 1958 out of the National Advisory Committee for Aeronautics (NACA) in a act of reorganization as an answer to the political pressure after the launch of Sputnik 1 by the Soviets. NASA started with a bulk of 8000 employees and an annual budget of roughly 100 Mio USD;
see Bilstein, 1989; Garber 1997; Ertel and Morse, 1969;
http://www.hq.nasa.gov/office/pao/History/sputnik/index.html, 2007-11-08.

[5] NASA officials contracted out between 80 and 90 percent of the funds they received each year, namely to contractors in Sunbelt states; Launius and McCurdy (1997): 241.